\documentclass[review]{elsarticle}

\usepackage{amssymb}
\usepackage{amsmath}
\usepackage{hyperref}
\usepackage{color}
\usepackage{float}
\usepackage{tikz}
\usepackage{graphicx}
\usepackage{subfigure}
\usepackage{pgfplots}
\usetikzlibrary{plotmarks}
\journal{Journal of Computational Physics}

\begin{document}

\begin{frontmatter}

\title{A weakly compressible SPH method for violent multi-phase flows with high density ratio}

\author{Massoud Rezavand}
\ead{massoud.rezavand@tum.de}

\author{Chi Zhang}
\ead{c.zhang@tum.de}

\author{Xiangyu Hu\corref{cor1}}
\ead{xiangyu.hu@tum.de}

\address{Department of Mechanical Engineering, 
	Technical University of Munich, 85748 Garching, Germany}

\cortext[cor1]{Corresponding author. Tel.: +49 89 289 16152}

\begin{abstract}
The weakly compressible SPH (WCSPH) method is known suffering from low computational efficiency, 
or unnatural voids and unrealistic phase separation 
when it is applied to simulate highly violent multi-phase flows 
with high density ratio, such as that between water and air. 
In this paper, to remedy these issues, we propose a multi-phase WCSPH method 
based on a low-dissipation Riemann solver and the transport-velocity formulation. 
The two-phase Riemann problem is first constructed to handle the pairwise interaction between fluid particles, 
then modified for the fluid-wall interaction to impose the solid wall boundary condition.
Since the method uses the same artificial speed of sound 
for both heavy and light phases, the computational efficiency increases greatly.
Furthermore, due to the transport-velocity formulation employed for the light phase 
and application of the two-phase Riemann problem,
the unnatural voids and unrealistic phase separation are effectively eliminated.
The method is validated with several 2- and 3D cases involving violent water-air flows.
The results have been compared with existing experimental data, previous numerical and analytical solutions, 
where the proposed method demonstrates good robustness, improved or comparable accuracy, respectively, 
comparing to previous methods with same choice of sound speed or those with much less computational efficiency. 
\end{abstract}


\end{frontmatter}


\section{Introduction}
Multi-phase flow phenomena, where different fluids are separated by phase interfaces, 
are ubiquitous in many natural and engineering problems. 
Due to the discontinuous properties and physical interactions 
across the dynamically evolving phase interface, 
the study of multi-phase flow problems is a challenging topic 
in computational fluid dynamics (CFD). 
Compared to the Eulerian mesh-based methods 
\cite{Hirt_1981_VoF,Sussman_1994_LSM,Tryggvason_2001_front,ABOULHASANZADEH2012456}, 
smoothed particle hydrodynamics (SPH), 
as a fully Lagrangian and mesh-free method, tracks the phase interfaces 
with no need for additional interface tracking or capturing algorithms, 
which are often suffered from serious numerical errors and instabilities. 
This advantage nominates SPH as a well suited 
candidate for the simulation of incompressible multi-phase flows and has attracted a great deal of 
attention in recent decades \cite{COLAGROSSI_Landrini_2003, Hu_Adams_2007_ISPH, 
	Zainali_2013_ISPH, Monaghan_Rafiee_2013_simple,KRIMI201853}.

The most widely used SPH method for incompressible multi-phase flows 
is the weakly compressible SPH (WCSPH) method, which was first proposed for simulating 
single-phase flows with and without free surfaces \cite{monaghan1994simulating, Morris_1997_Low_Reynolds}.  
In the WCSPH method, the fluid pressure is related to density using an artificial equation of state 
allowing a small compressibility. 
When the flow is violent, e.g. inviscid or with high Reynolds number, 
the standard WCSPH formulation can produce spurious pressure oscillations, 
thereby may lead to numerical instabilities. 
To alleviate this issue, several methods, 
including artificial viscosity \cite{monaghan1983shock}, 
density filtering \cite{COLAGROSSI_Landrini_2003} and those based on Riemann solver 
\cite{vila1999particle,moussa2006convergence,antuono2010free}, have been proposed. 

As the issue of numerical instabilities become even more serious in the simulation of 
violent multi-phase flows, especially for those with large density ratio, 
such as that of water and air,    
Colagorossi and Landrini \cite{COLAGROSSI_Landrini_2003} 
further proposed several combined modifications, 
including Mean Least Square (MLS) 
interpolation to frequently reinitialize the density field, 
gradient discretization based on volume (not density) to remedy the density discontinuity issues 
at the interface, a cohesion term for air to artificial surface-tension effects and 
XSPH formulation, which corrects the velocity term used in 
the density and position evolution equations, 
to regularize particle distribution of the heavy phase. 
These modifications to stabilize computations 
not only increase the complexity of the original simple WCSPH formulation, 
but also their effectiveness relies on a computationally very expensive approach, 
i.e. assigning the light phase with an artificial speed of sound typically 10 times 
larger than the heavy phase to obtain a reasonable pressure profile across the interface. 
Following Colagorossi and Landrini, 
Mokos et al. \cite{Mokos_2015_Multiphase} 
have carried out high resolution simulations of violent water-air flows. 
The simulation results exhibit non-physical voids and phase separation. 
Such artifacts have also been observed in Gong et al. \cite{gong2016two}. 
To avoid unnatural voids in such simulations 
a regularization method based on particle shifting \cite{Xu_2009_Shifiting} was introduced \cite{mokos2017shifting}. 
A further improvement for simulating high Reynolds number flows is also proposed 
by applying a density diffusive terms \cite{antuono2010free} for smoother pressure and density fields \cite{hammani2018}. 
However, the very large speed of sound for the light phase 
is still required to produce acceptable simulation results.

Different from Colagorossi and Landrini \cite{COLAGROSSI_Landrini_2003}, 
Chen et al. \cite{chen2015multi} proposed another set of combined modifications, 
including pressure gradient operators 
imposing continuously induced particle acceleration across the interface \cite{hu2007incompressible}, 
a positive background pressure to avoid negative pressure 
and a similar density reinitialization to obtain a smooth pressure field. 
Since the same speed of sounds are used for the dense and light fluids, 
the computational efficiency increases greatly. 
A considerable drawback of this method is the excessive numerical dissipation, 
which is able to stabilize simulations but also 
leads to large discrepancies between their and the reference simulation results, 
such as those in Refs. \cite{COLAGROSSI_Landrini_2003, Mokos_2015_Multiphase}, 
especially when the density ratio is high.
Zheng and Chen \cite{ZHENG2019177} proposed a further modification 
with a first-order density reinitialization 
and a local implementation of the artificial viscosity.
This modification demonstrates enhanced accuracy, 
such as shorter delays of the dynamics, due to less numerical dissipation. 
However, as shown in their only violent water-air flow case, 
the particles still undergo spurious fragmentation, particularly in the vicinity of phase interface.

In the present study, we propose a simple WCSPH method to simulate highly 
violent multi-phase, typically liquid-gas, flows with wave breaking and impact events. 
Our method is based on two modifications of the original WCSPH formulation:
one is a two-phase extension from the low-dissipation Riemann solver \cite{zhang2017weakly} 
to realize particle interactions and to stabilize the interface interaction;
the other is applying the transport-velocity formulation \cite{Hu_Adams_2006_WCSPH} 
in the light phase to eliminate non-physical voids or fragmentations.
As a result of the two-phase Riemann solver, 
the light phase experiences the heavy phase like moving wall boundary, 
while the heavy phase undergoes a free-surface-like flow with variable free-surface pressure. 
Since we use the same value for the artificial speed of sound in both light and heavy phases, 
the computational efficiency is much higher than 
those in Refs. \cite{COLAGROSSI_Landrini_2003,Monaghan_Rafiee_2013_simple,hammani2018}. 
The low-dissipation nature of the proposed method accurately captures 
the violent impacts and breaking events, without suffering from 
excessive numerical dissipation \cite{chen2015multi}
or spurious fragmentation \cite{ZHENG2019177}. 
Note that, we do not use other stabilizing methods to tackle the density discontinuity at the 
phase interface \cite{Khayyer2013_HighDensityRatio,Zainali_2013_ISPH,rezavand2018isph}, 
nor the artificial repulsive pressure force \cite{Monaghan_Rafiee_2013_simple,rezavand2018isph}. 
Furthermore, density reinitialization schemes, deemed to be necessary in the literature 
\cite{COLAGROSSI_Landrini_2003,chen2015multi,gong2016two,ZHENG2019177}, are not required. 
In order to test the accuracy and robustness of the presented method, 
several 2D and 3D benchmark tests involving highly violent water-air flows with free 
surfaces are carried out and the results are validated against existing 
experimental data, previous numerical simulations and analytical solutions.
The paper is organized as follows: Section \ref{sec:NumericalModel} details the proposed multi-phase scheme. 
The numerical results are then presented 
and discussed in Section \ref{sec:ResultsandDiscussion}. Finally, the 
concluding remarks of the present study are given in Section \ref{sec:conclusions}.

\section{Numerical method}\label{sec:NumericalModel}
For an inviscid and immiscible two-phase, typically liquid-gas, flow with large density ratio,
i.e. $ \rho_l/\rho_g \gg 1$, where $\rho_l$ and $\rho_g$ 
represent the densities of heavy and light phases,    
the mass and momentum conservation equations can be 
written respectively as
\begin{equation}\label{eq:massGeneral}
\frac{d\rho}{dt} = -\rho\nabla\cdot\mathbf{v},
\end{equation}
\begin{equation} \label{eq:momentumGeneral}
\frac{d\mathbf{v}}{dt} = -\frac{1}{\rho}\nabla p + \mathbf{g},
\end{equation}
where $d/dt$ is the material or Lagrangian derivative, $\rho$ is the density, 
$\mathbf{v}$ is the velocity, $p$ is the pressure and $\mathbf{g}$ is 
the gravitational acceleration. 
To close the system, pressure is estimated from density via an artificial equation of state, 
within the weakly compressible regime. Here, 
we use a simple linear equation for both the heavy and light phases
\begin{equation}\label{eq:EoS}
p = c^{2}(\rho-\rho_0),
\end{equation} 
where $c$ and $\rho_0$ are the speed of sound and the initial reference density. 
Here, we assume that the speed of sound is constant 
and we set $c=10U_{max}$, where $U_{max}$ denotes the 
maximum anticipated velocity of the flow. Note that, 
different from Refs. \cite{COLAGROSSI_Landrini_2003, 
	Monaghan_Rafiee_2013_simple,Mokos_2015_Multiphase}, 
we use the same speed of sound for both heavy and light phases
as in Refs. \cite{chen2015multi, ZHENG2019177}.

In the WCSPH formulation, there are two different formulations to implement mass conservation, 
viz. particle summation and continuity equation \cite{monaghan2005,Monaghan_2012_Annual_Rew}. 
The former calculates density through a summation over all the neighboring particles
\begin{equation}\label{eq:densitySummation}
\rho_i = m_i\sum_{j}W_{ij},
\end{equation}
where $m_i$ denotes the mass of particle $i$ and the smoothing kernel function 
$ W(\left| \mathbf{r}_{ij}\right| ,h)$ is simply substituted by $ W_{ij} $, 
with $\mathbf{r}_{ij}=\mathbf{r}_{i}-\mathbf{r}_{j}$ and $h$ being the 
displacement vector between particle $i$ and $j$ and the smoothing length, 
respectively.
The latter updates particle density by discretizing the continuity equation
\begin{equation} \label{eq:SPH_massConservation}
\frac{d\rho_{i}}{dt} = \rho_i \sum_{j}\frac{m_j}{\rho_j}\mathbf{v}_{ij}\cdot\nabla_{i} 
W_{ij}=2\rho_i\sum_{j} \frac{m_j}{\rho_j} (\mathbf{v}_{i}- \mathbf{\overline{u}}_{ij}) 
\cdot\nabla_{i} W_{ij},
\end{equation}
where $\mathbf{v}_{ij}=\mathbf{v}_{i}-\mathbf{v}_{j}$ is the relative velocity 
and $\mathbf{\overline{u}}_{ij}=(\mathbf{v}_{i}+\mathbf{v}_{j})/2$ 
the average velocity between particle $i$ and $j$.
 
Following \cite{monaghan2005,Monaghan_2012_Annual_Rew},
without taking the artificial viscosity into account, 
the momentum conservation equation can be discretized as
\begin{equation} \label{eq:SPH_momentumConservation}
\frac{d\mathbf{v}_{i}}{dt} = -\sum_{j}m_j \left( \frac{p_i + p_j}{\rho_i\rho_j}\right)
\nabla_{i} W_{ij}=-2\sum_{j} m_j \frac{\overline{{p}}_{ij}}{\rho_i\rho_j} \nabla_{i} W_{ij},
\end{equation}  
with $\overline{{p}}_{ij}=(p_{i}+p_{j})/2$ being the average pressure between 
particle $i$ and $j$. 

\subsection{WCSPH based on a multi-phase Riemann solver}\label{subsec:multiphase_RiemannbasedSPH}
\begin{figure}\centering
	\subfigure []{\includegraphics[width=0.35\linewidth]{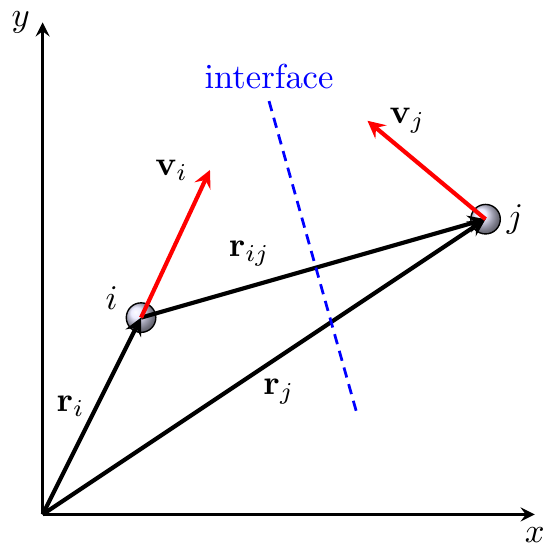}} 
	\subfigure []{\raisebox{-4pt}{\includegraphics[width=0.55\linewidth]{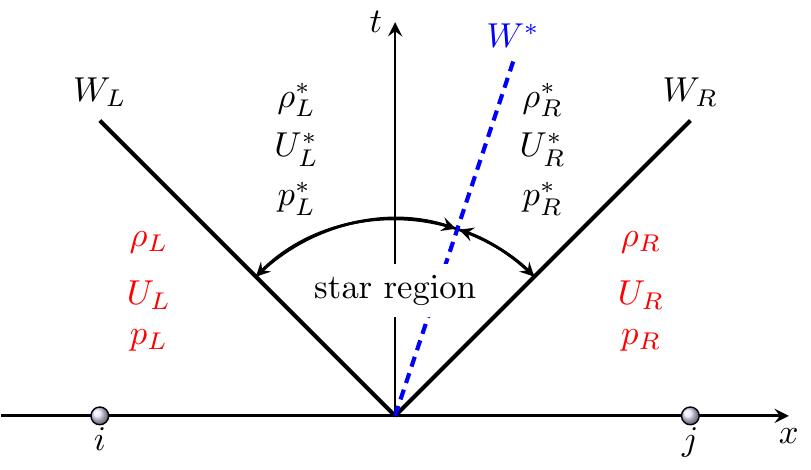}}}
	\caption{Multi-phase Riemann problem: 
		(a) Construction along the interface between particle $i$ and $j$.
	(b) Structure of the solution.  }
	\label{fig:RS}
\end{figure}

For an SPH methods based on Riemann solver~\cite{vila1999particle,moussa2006convergence}, 
an inter-particle Riemann problem is constructed 
along the unit vector $\mathbf{e}_{ij}=-\frac{\mathbf{r}_{ij}}{\left| \mathbf{r}_{ij}\right|}$ 
as shown in Fig. \ref{fig:RS}(a), with the initial left and right states reconstructed 
from particle $i$ and $j$ by piecewise constant approximation, respectively, as follows
\begin{equation}\label{eq:LRstates}
\begin{cases}
(\rho_L, U_L, p_L) = (\rho_i, \mathbf{v}_i \cdot \mathbf{e}_{ij}, p_i),\\
(\rho_R, U_R, p_R) = (\rho_j, \mathbf{v}_j \cdot \mathbf{e}_{ij}, p_j),
\end{cases}
\end{equation}
where subscripts $L$ and $R$ denote left and right states, respectively, and the 
discontinuity is assumed at $\mathbf{\overline{r}}_{ij}=(\mathbf{r}_{i}+\mathbf{r}_{j})/2$.
The solution of the Riemann problem can be shown using $x-t$ diagrams \cite{toro1989fast}, 
as in Fig. \ref{fig:RS}(b). Three waves are present: $W_L$,$W^*$ and $W_R$ 
respectively at the left, middle and right. The left and right waves are either a shock 
or a rarefaction, while the middle wave is always a contact discontinuity separating 
the two intermediate states, namely, ($\rho^*_L, U^*_L, p^*_L$) and ($\rho^*_R, U^*_R, p^*_R$). 
The intermediate states are regarded to satisfy the interface condition $p^*=p^*_L = p^*_R$ and $U^*=U^*_L = U^*_R$. 
For multi-phase flows with the same speed of sound across the interface, 
following Ref. \cite{hu2004interface}, the intermediate velocity and  pressure can be approximated as 
\begin{equation}\label{eq:RSsolution}
\begin{cases}
U^* = \overline{U} +\frac{p_L - p_R}{c(\rho_L+\rho_R)},\\
p^* = \overline{P} + \frac{\rho_L\rho_R\beta(U_L - U_R)}{\rho_L+\rho_R},
\end{cases}
\end{equation}
where $\overline{U} = (\rho_L U_L + \rho_R U_R)/(\rho_L+\rho_R)$ 
and $\overline{P} = (\rho_L p_R + \rho_R p_L)/(\rho_L+\rho_R)$.
In Eq. (\ref{eq:RSsolution}), 
each solution for velocity and pressure consists of two terms.
The first term is a density-weighted average 
and the second term is corresponding to numerical dissipation.
Due to the assumption of weak compressibility, 
the magnitudes of the second term (numerical dissipation) are much smaller than that of the first term. 
Note that, following Zhang et al. \cite{zhang2017weakly}, 
a limiter $\beta= \mathrm{min}[3 \mathrm{max} (U_L - U_R,0), c]$ 
is applied for $p^*$ in Eq. \eqref{eq:RSsolution} to reduce the numerical dissipation. 
Also note that, for a single phase interaction, 
Eq. (\ref{eq:RSsolution}) recovers the solution of 
the low-dissipation Riemann solver \cite{zhang2017weakly}.   
For two-phase particle interactions with large density ratio,
Eq. (\ref{eq:RSsolution}) gives approximately 
the intermediate velocity from the heavy phase, 
which indicates that the light phase experiences the heavy phase as a moving wall boundary, 
and intermediate pressure from the light phase, 
which indicates that the heavy phase undergoes a free-surface-like 
flow with variable free-surface pressure \cite{hu2004interface}.     
Having the intermediate values determined, the mass and  momentum conservation equations, 
i.e. Eqs. \eqref{eq:SPH_massConservation} and \eqref{eq:SPH_momentumConservation}, 
can be rewritten as
\begin{equation} \label{eq:SPH_star_massConservation}
\frac{\text{d}\rho_{i}}{\text{d}t} = 2\rho_i\sum_{j} \frac{m_j}{\rho_j} (\mathbf{v}_{i}- \mathbf{v}^{*}) \cdot\nabla_{i} W_{ij},
\end{equation}
\begin{equation} \label{eq:SPH_star_momentumConservation}
\frac{\text{d}\mathbf{v}_{i}}{\text{d}t} = -2 \sum_{j} m_j \frac{{p}^{*}}{\rho_i \rho_j} \nabla_{i} W_{ij},
\end{equation}
where $\mathbf{v}^{*} = U^{*} \mathbf{e}_{ij} 
+ (\overline{\mathbf{v}}_{ij} - \overline{U}\mathbf{e}_{ij} )$. 
Here, for consistency, one can assign $\overline{\mathbf{v}}_{ij} 
= (\rho_i \mathbf{v}_i + \rho_j \mathbf{v}_j)/(\rho_i+\rho_j)$,
which is a density-weighted average velocity between particle $i$ and $j$. 

\subsection{Transport velocity formulation}
Due to the tensile instability, WCSPH results in particle clumping and unnatural 
void regions in fluid dynamics problems \cite{monaghan2000sphTensile}. In particular, 
negative pressure gives rise to particle clustering. 
Such artifacts become even more serious in multi-phase SPH simulations of water-air flows 
\cite{COLAGROSSI_Landrini_2003,mokos2017shifting}. 
Colagrossi and Landrini \cite{COLAGROSSI_Landrini_2003} applied a background pressure 
to the momentum conservation equation to avoid the tensile instability, 
while Mokos et al. \cite{mokos2017shifting} 
additionally proposed a particle shifting method to prevent particle clumping.

Here, to prevent the unnatural void regions observed in the light phase, 
we apply the transport-velocity formulation \cite{adami2013transport, zhang2017generalized} 
on the particles of light phase, 
but not on those of the heavy phase because the pressure on latter generally 
is kept positive due to the gravity.
In transport-velocity formulation, the particle advection velocity  $\tilde{\mathbf{v}}$, 
which is used to advect the particles of light phase, 
is obtained by using a constant background pressure $p_b$ at every time step of size $\delta t$
\begin{equation} \label{eq:transprort_vel}
\tilde{\mathbf{v}}_i(t + \delta t) = \mathbf{v}_i(t)  + \delta t \left( \frac{d\mathbf{v}_i}{dt} 
- 2p_b\sum_{j} m_j\frac{1}{\rho_i \rho_j} \nabla_{i} W_{ij}\right).
\end{equation}
Note that, at the interface, 
water particles are also included in the summation of Eq. \eqref{eq:transprort_vel}.
With the  transport-velocity formulation, the momentum conservation equation for light phase 
is discretized as  
\begin{equation} \label{eq:SPH_star_momentum_air}
\frac{d\mathbf{v}_{i}}{dt} = -2\sum_{j} m_j \frac{{p}^{*}}{\rho_i\rho_j} \nabla_{i} W_{ij} 
+ 2 \sum_{j} m_j\frac{\overline{\mathbf{A}}_{ij}}{\rho_i \rho_j}
\cdot \nabla_{i} W_{ij} + \mathbf{g}_i,
\end{equation}
where $\overline{\mathbf{A}}_{ij} = (\mathbf{A}_i + \mathbf{A}_j)/2$. 
Here, $\mathbf{A}_i=\rho_i \mathbf{v}_i(\tilde{\mathbf{v}}_i-\mathbf{v}_i)$ 
is an extra stress tensor as a consequence of advection velocity $\tilde{\mathbf{v}}_i$.
Note that at the phase interface $\mathbf{A}_j$ vanishes for the particles of heavy phase 
because the transport-velocity formulation is not applied, 
i.e. $\tilde{\mathbf{v}}_j = \mathbf{v}_j$. 
Also note that, following transport-velocity formulation,
we employ the density summation form of Eq.~\eqref{eq:densitySummation} for the light phase.

\subsection{Wall boundary condition}
In the present study, we use fixed dummy particles to impose the solid wall 
condition as proposed in Ref. \cite{ADAMI2012wall}. 
To realize the fluid-wall interactions 
a Riemann problem is constructed between particles of fluids and wall dummy particles, 
as for fluid-fluid interactions (see Section \ref{subsec:multiphase_RiemannbasedSPH}). 
However, the intermediate pressure value is obtained as
\begin{equation}\label{eq:RSatWall}
p^* = \frac{\rho_f p_w + \rho_w p_f}{\rho_f+\rho_w},
\end{equation}
where subscripts $f$ and $w$ denote fluid and wall, respectively, 
to decrease the wall-induced numerical dissipation. 
Note that the intermediate velocity value $U^*$ is still obtained via Eq. \eqref{eq:RSsolution}, 
as for fluid-fluid interactions. 
Similar to Ref. \cite{ADAMI2012wall}, the pressure of 
wall dummy particles is calculated by the summation over all contributions of the 
neighboring fluid particles as
\begin{equation}\label{eq:wallPressure}
p_w = \frac{\sum_{f}\frac{p_f}{\rho_f} W_{wf} + \mathbf{g}\sum_{f} \mathbf{r}_{wf} 
	W_{wf}}{\sum_{f}\frac{W_{wf}}{\rho_f}}.
\end{equation}
Note that, by introducing $\rho_f$ into the above equation
the contribution of heavy phase particles in $p_w$ at a triple point, 
where water, air and solid particles meet, is vanishing.
Again, following Ref. \cite{ADAMI2012wall}, 
the density of wall dummy particles is obtained form pressure via Eq. (\ref{eq:EoS}).

\subsection{Time integration}
To integrate the equations of motion in time, the kick-drift-kick 
\cite{monaghan2005,adami2013transport} scheme is employed. 
The first half-step velocity is obtained as
\begin{equation}
\mathbf{v}_i^{n + \frac{1}{2}} = \mathbf{v}_i^n +  \frac{\delta t}{2}  \big( \frac{d \mathbf{v}_i}{dt} \big)^{n},
\end{equation}
from which we obtain the position of particles at the next time step
\begin{equation}
\mathbf{r}_i^{n + 1} = \mathbf{r}_i^n +  \delta t  \mathbf{v}_i^{n + \frac{1}{2}}.
\end{equation}
With the updated flow states, the density of heavy-phase particles at the new 
time step is then calculated as
\begin{equation}\label{eq:newRho}
\rho_i^{n + 1} = \rho_i^n +  \delta t  \big( \frac{d \rho_i}{dt} \big)^{n+1},
\end{equation} 
where the time increment of density is approximated via Eq. \eqref{eq:SPH_massConservation}. 
Note that, the density of light phase particles at the new time step 
is obtained using the density summation equation, i.e. Eq. \eqref{eq:densitySummation}. 
Having new densities calculated, 
the pressure of particles and the time increment of velocity, can be obtained. 
Finally, the velocity of particles is updated for the new time step as
\begin{equation}
\mathbf{v}_i^{n + 1} = \mathbf{v}_i^{n + \frac{1}{2}} +  \frac{\delta t}{2}  
\big( \frac{d \mathbf{v}_i}{dt} \big)^{n+1}.
\end{equation}
For numerical stability, the time-step size is limited by the CFL condition
\begin{equation}
\delta t \leq 0.25 \big( \frac{h}{c + U_{max}}  \big),
\end{equation}
and the body force condition
\begin{equation}
\delta t \leq 0.25 \sqrt \frac{h}{\left|\mathbf{g} \right| }.
\end{equation}

\section{Numerical results}\label{sec:ResultsandDiscussion}
In this section several test cases involving two-phase dam break and sloshing
are considered to validate the proposed method 
for modeling violent flows with high density ratio. 
The 5th-order Wendland kernel \cite{Wendland1995}  
with a smoothing length of $h = 1.3 dx$, 
where $dx$ is the initial particle spacing, 
and a support radius of $2.6dx$ are used. 
In all cases, the maximum velocity is estimated as $U_{max} = 2\sqrt{gH}$, 
where $H$ is the initial water depth, 
according to the shallow-water theory \cite{ritter1892fortpflanzung}, 
for setting the speed of sound $c_0$. 
Here, we set the background pressure $p_b = 4\rho_0 c^2_0$ 
for the transport-velocity formulation in all of the test cases.

\subsection{Two-phase dam break}\label{subsec:twophase_dambreak}
We consider a two-phase dam break case, which has been experimentally 
studied by Zhou et al. \cite{zhou1999nonlinear}. 
Owing to its complex flow involving wave impact and breaking events, 
dam break is widely used to validate SPH-based methods 
\cite{COLAGROSSI_Landrini_2003,ADAMI2012wall,LEROY2014unified,
Mokos_2015_Multiphase,chen2015multi}. 
Note that, in the two-phase model, 
the cushioning effect of entrapped air pockets during violent hydrodynamics can influence the impact loads. 

\begin{figure}[tb!]
	\centering
	\includegraphics[width=0.5\linewidth]{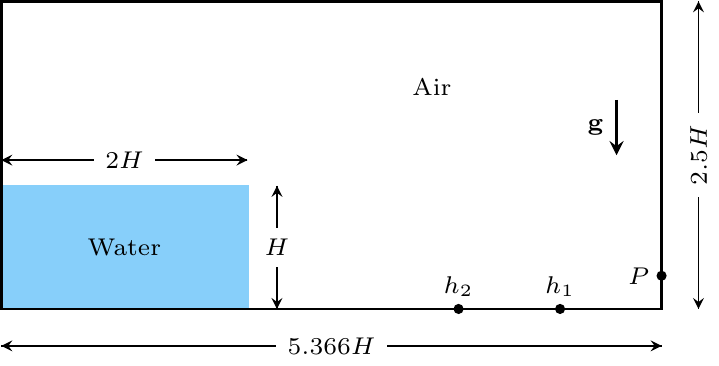}
	\caption{Schematic illustration of the two-phase dam break problem.}
	\label{fig:2phase_dambreak_config}
\end{figure}

The schematic of this problem is illustrated in Fig. \ref{fig:2phase_dambreak_config}. 
A water column of height $H=1$ and width of $2H$ is located in a tank of length $5.366H$, 
and is surrounded by air. 
We consider an inviscid flow, where the density of water 
and air are set to $\rho_w = 1$ and $\rho_a = 0.001$, respectively. 
Using $g=1$ in negative $y$-direction as the gravity, 
all quantities correspond to their non-dimensional variables.
Note that for simplicity, water with zero initial pressure is released 
immediately when the computation starts 
instead of being released from a gate holding the pressure-relaxed water as in the experimental setup.
To record the pressure signals, a sensor $P$ is located on the downstream wall at 
$y/H=0.19$ from the bottom wall, which is slightly different from the experiments. 
Note that this adjustment gives better agreement between SPH and 
experimental results due to several uncertainties in the measurements \cite{greco2001two}. 
In the simulation, the measured pressure is obtained by averaging the values from particles 
within a support radius $2.6dx$ with 
\begin{equation}\label{eq:sensorpressure}
P =  \frac{\sum p_f W_{sf} V_f}{\sum W_{sf} V_f + \epsilon } ,
\end{equation}
where subscripts $s$ and $f$ denote solid wall and fluids, respectively, 
$V_f = m_f/\rho_f$ is the volume of a fluid particle
and $\epsilon = 1.0 \times 10^{-15}$ is a small positive number to avoid dividing by zero. 
Furthermore, two height sensors $h_1$ and $h_2$ are located on the bottom bed at 
$x_1/H=0.825$ and $x_2/H=1.653$ from the right wall, respectively, to monitor the 
time evolution of the water level. 

\begin{figure}[tb!]
	\centering
	\includegraphics[width=\linewidth]{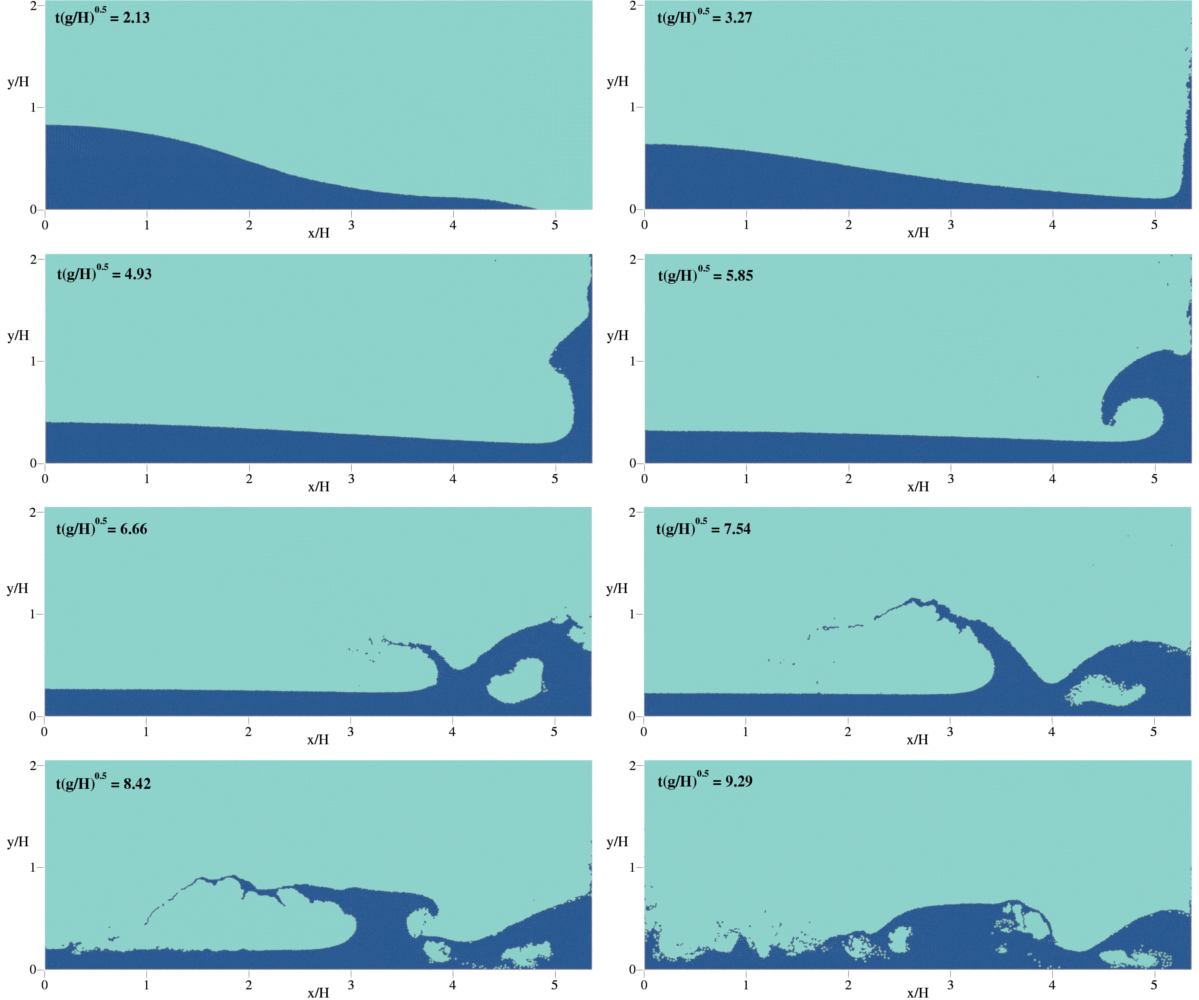}
	\caption{Two-phase dam break: snapshots of the computational results 
		at eight different time instants 
		$t\sqrt{g/H}=$2.13, 3.27, 4.93, 5.85, 6.66, 7.54, 8.42 and 9.29 with $H/dx=80$.}
	\label{fig:dambreak_8frames}
\end{figure}

Fig. \ref{fig:dambreak_8frames} 
illustrates several snapshots during the time evolution.  
It can be observed that the main features of the flow, viz. 
high roll-up along the downstream wall, a large reflected jet and a water 
surface breaking due to the re-entry of the backward wave, are well captured by the present method, 
as in previous single-fluid  numerical simulations (see e.g. 
Refs. \cite{ferrari2009new, zhang2017weakly, Zhang_efficientSPH_2018}). 
Moreover, the water-air interface is sharply maintained 
as in previous two-phase simulations (see e.g. Refs. \cite{COLAGROSSI_Landrini_2003,marrone2016analysis}). 
Note that, due to the transport-velocity formulation applied in air phase 
and the application of two-phase Riemann problem, 
no obvious unnatural voids or unrealistic phase separation appears.

\begin{figure}\centering
	\subfigure []{\includegraphics[width=0.25\linewidth]{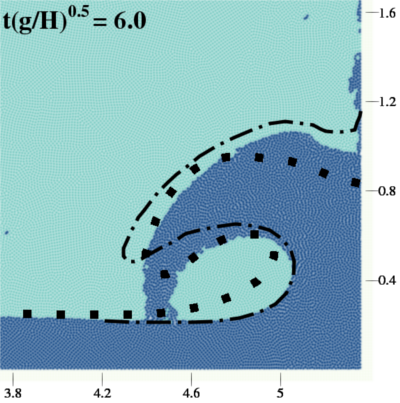}} 
	\subfigure []{\includegraphics[width=0.25\linewidth]{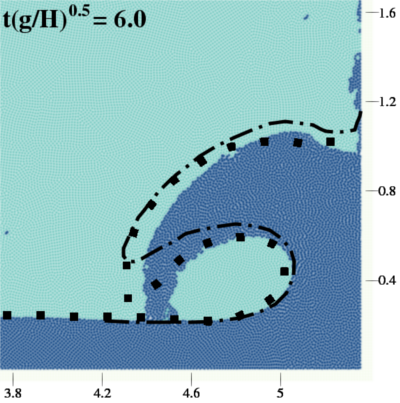}}
	\subfigure []{\raisebox{1pt}{\includegraphics[width=0.475\linewidth]{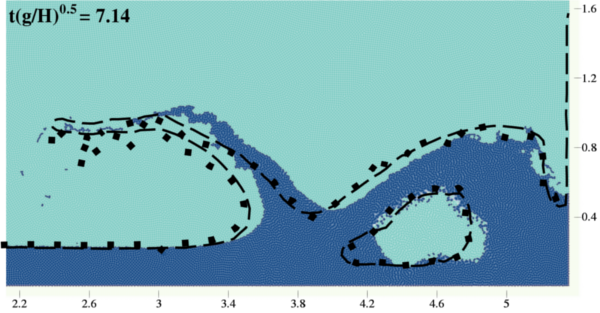}}}
	\caption{Two-phase dam break: evolution of the water-air interface 
				obtained by the present SPH method with $H/dx=80$, in comparison with those obtained by (a) 
				the BEM method in Ref. \cite{ANTUONO20122570} (dash-dotted lines) 
				and the two-phase SPH method in Chen et al. \cite{chen2015multi} ($\blacksquare$), 
				(b) the same BEM method (dash-dotted lines) and the two-phase SPH method
				in Zheng and Chen \cite{ZHENG2019177} ($\blacksquare$), and (c) 
				a two-phase level-set method \cite{colicchio2005level} (dashed lines) 
				and the two-phase SPH method in Colagrossi and Landrini \cite{COLAGROSSI_Landrini_2003} 
				($\blacksquare$).}
	\label{fig:dambreak_SPHvsLS}
\end{figure}

To evaluate the accuracy of the captured phase interface, 
as shown in Fig. \ref{fig:dambreak_SPHvsLS}, 
the present results are compared with those obtained by the boundary element method (BEM)
in Ref. \cite{ANTUONO20122570}, the level-set method in Refs.\cite{colicchio2005level, COLAGROSSI_Landrini_2003}, 
and previous two-phase SPH methods, 
namely, Refs. \cite{COLAGROSSI_Landrini_2003}, \cite{chen2015multi} and \cite{ZHENG2019177}.
All the three previous SPH simulations are with a spatial resolution of $H/dx=60$, 
which is comparable to the spatial resolution of our simulation, i.e. $H/dx=80$.
As it can be observed, a good agreement is achieved between the predicted interface location
by the presented method herein, 
and the BEM and level-set predictions for the smooth part of the solutions.
However, due to the remarkable physical uncertainties of the breaking waves, 
there exist discrepancies. 
On the other hand, as it is also observed in Ref. \cite{ANTUONO20122570}, 
the BEM method leads to excessive numerical dissipation 
and its results show considerable time delay. 

The following comparison is noteworthy between the present method and the two 
classes of the previous SPH simulations, viz. Refs. \cite{COLAGROSSI_Landrini_2003} 
and \cite{chen2015multi,ZHENG2019177}: 
one can observe that the present method accurately predicts 
a sharply maintained phase interface (Fig. \ref{fig:dambreak_SPHvsLS}(c)), 
while is computationally much more efficient than the method presented in Ref. \cite{COLAGROSSI_Landrini_2003}, 
due to algorithm simplicity and the same speed of sound for both the heavy and light phases, 
while there exist no unnatural void region as observed in Refs. \cite{chen2015multi,ZHENG2019177} 
(see Figs. \ref{fig:dambreak_SPHvsLS}(a) and (b)). 
Albeit the methods presented in Refs. \cite{chen2015multi,ZHENG2019177} are computationally efficient 
(also due to the same speed of sound for both the heavy and light phases), 
they suffer from spurious fragmentation at the phase interface, as well as non-physical void regions.
As can be seen in Figs. \ref{fig:dambreak_SPHvsLS}(a) and (b), 
the excessive numerical dissipation exhibited in Ref. \cite{chen2015multi} 
(using a global artificial viscosity) is, to some degree, 
alleviated by the method presented in Ref. \cite{ZHENG2019177}
(using a local artificial viscosity), however, 
the phase interface undergoes more severe fragmentations in the latter 
(see the figures in their Section 4.5).

\begin{figure}[H]
	\centering
	\includegraphics[width=0.75\textwidth]{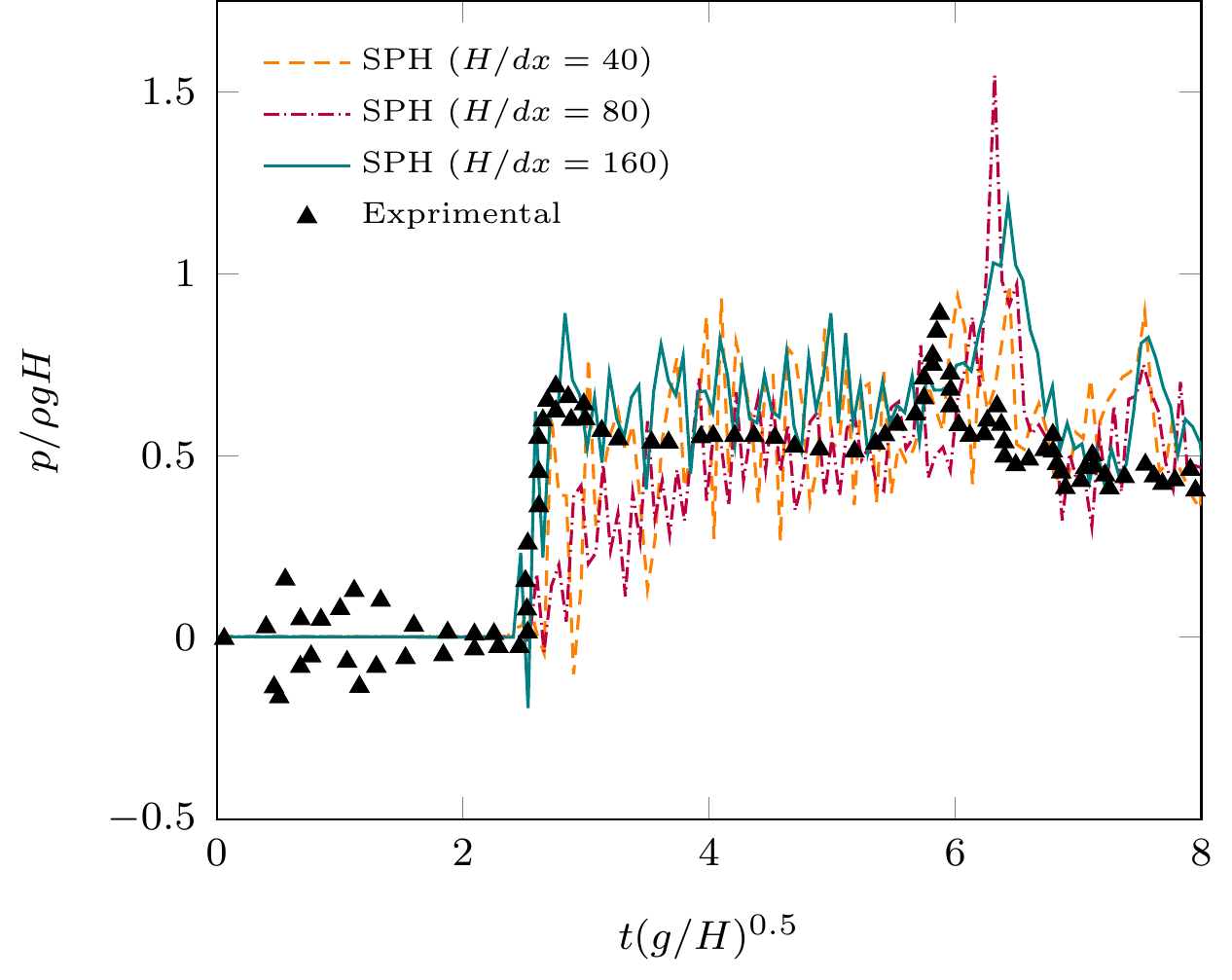}
	\caption{Two-phase dam break: comparison of the pressure signals recorded at 
		sensor $P$ between SPH predictions and experimental data \cite{zhou1999nonlinear}.}
	\label{fig:dam_pressure-plot}
\end{figure}

\begin{figure}
	\centering
	\includegraphics[width=0.75\textwidth]{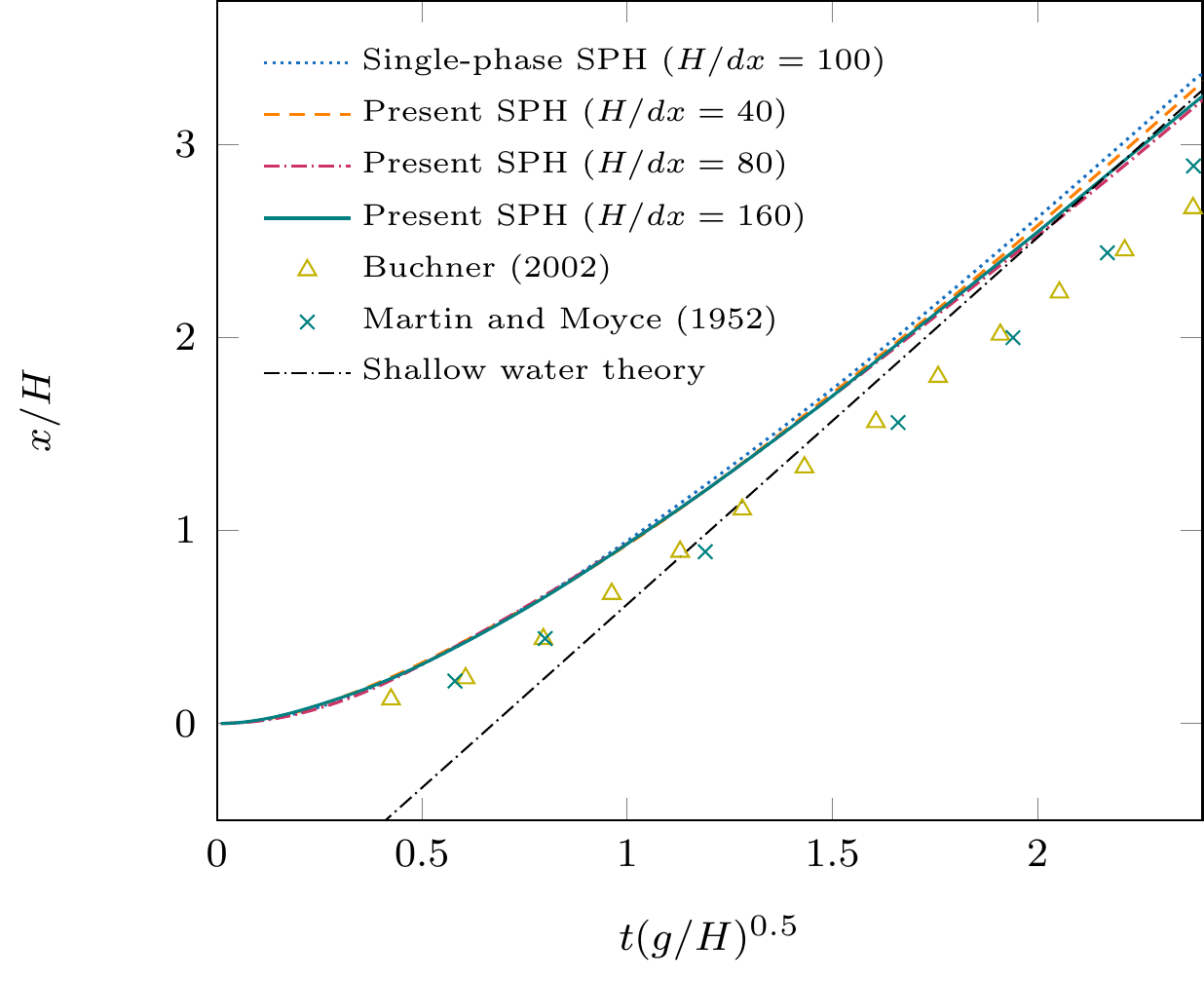}
	\caption{Two-phase dam break: convergence study on time history of 
		the water wave front in comparison with experimental data 
		\cite{buchner2002phd,martin1952part}, an analytical solution 
		derived from the shallow water theory \cite{ritter1892fortpflanzung},
		and a single-phase SPH simulation \cite{zhang2017weakly}.}
	\label{fig:dambreak_wavefront}
\end{figure}

Time history of the recorded impact pressure signals on the vertical downstream 
wall are compared against experimental data in Fig. \ref{fig:dam_pressure-plot}.
As expected, in a weakly compressible framework, the pressure profile exhibits 
high frequency oscillations, which are mitigated with increased spatial resolution.
With the fairly low resolution in mind, a good agreement in the overall behavior 
of the pressure is noted.  
The predicted water wave front with increasing particle resolution 
is compared against the experimental data provided by 
Refs. \cite{buchner2002phd} and \cite{martin1952part} in Fig. \ref{fig:dambreak_wavefront}, 
as well as the analytical solution from shallow water theory \cite{ritter1892fortpflanzung}.
The results are also compared with a previous single-phase SPH simulation \cite{zhang2017weakly}.
Due to the consideration of air, compared to the previous single-phase SPH results, 
the present two-phase simulation predicts slightly slower wave front
at time instants after $t\sqrt{g/H}\approx 2.0$, 
which suggest a closer agreement with experimental data. 
However, as observed in several previous works \cite{COLAGROSSI_Landrini_2003, ferrari2009new, MARRONE2011delta},
a slower front wave propagation is observed in experiments, which is attributed to 
several uncertainties, viz. accuracy of the experiments, turbulent wall boundary layer and wall roughness.
The asymptotic convergence of the predicted water wave front to the theoretical 
shallow water solution with increased spatial resolution can also be noted on a long time scale.
Note that the shallow water assumption does not hold at initial time instants of the problem.

\begin{figure}[tb!]
	\centering
	\begin{minipage}{.5\linewidth}
		\centering
		\includegraphics[height=4.8cm]
		{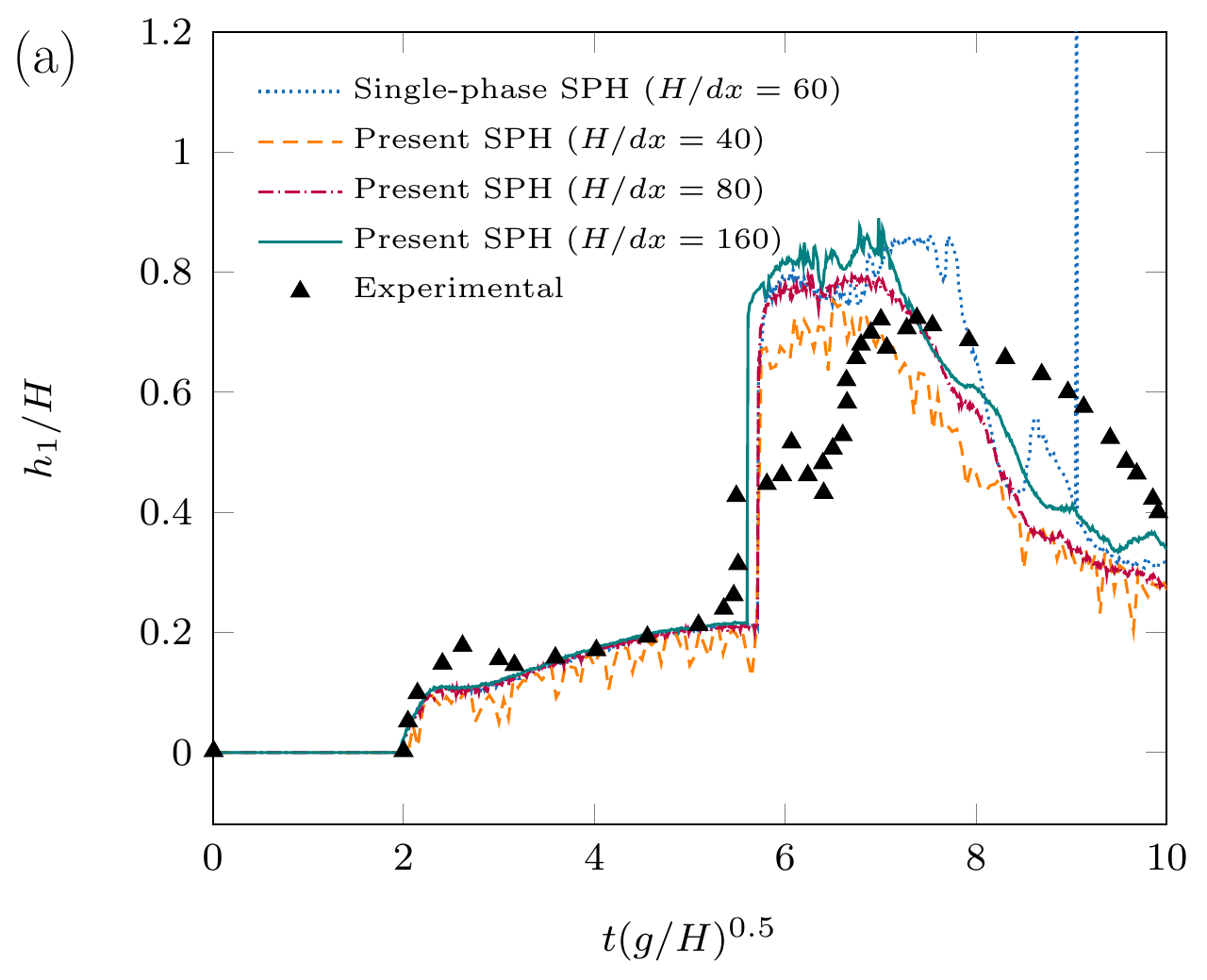}
	\end{minipage}%
	\begin{minipage}{.5\linewidth}
		\includegraphics[height=4.8cm]
		{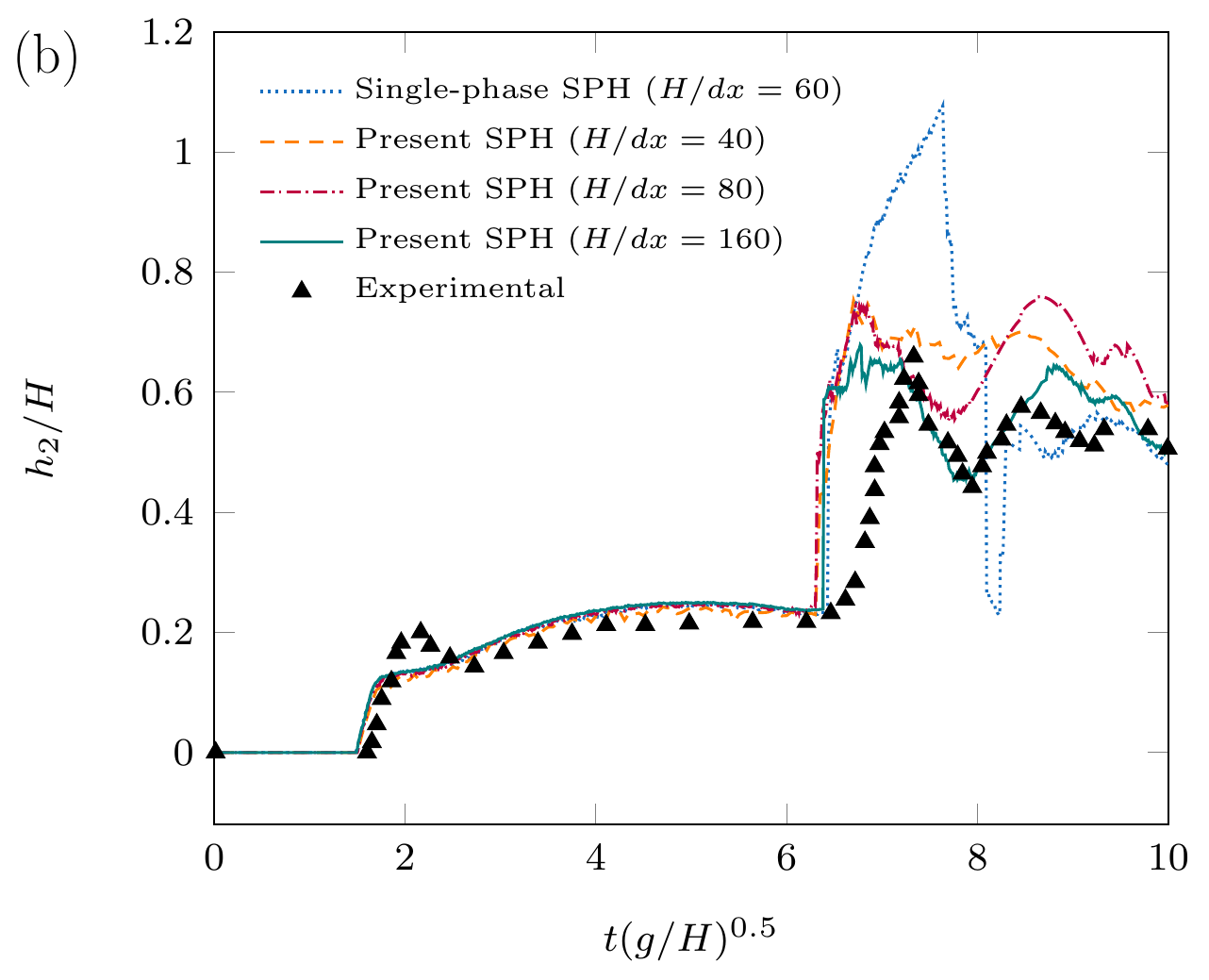}
	\end{minipage}
	\caption{Time history of the water surface level at (a) sensor $h_1$ and
		(b) sensor $h_2$, in comparison with experimental data \cite{zhou1999nonlinear}
		and single-phase SPH predictions presented in \cite{zhang2017weakly}.}
	\label{fig:dambreak_h1_h2}
\end{figure}

Fig. \ref{fig:dambreak_h1_h2} plots the time history of the recorded 
water level signals at sensors $h_1$ and $h_2$ in comparison with 
experimental data and single-phase SPH predictions.
As mentioned in Ref. \cite{COLAGROSSI_Landrini_2003}, 
since time evolution of the water level is determined by the wave front shape, 
initial conditions and the bottom bed roughness are influential. 
The differences between simulation results and the experiments are thus reasonable 
due to the present inviscid model and different initial conditions. 
Similar discrepancies are also observed in previous numerical studies 
\cite{COLAGROSSI_Landrini_2003,ferrari2009new}. 
Due to the slightly faster wave front predicted by the inviscid simulation 
(see Fig. \ref{fig:dambreak_wavefront}), higher water levels are recorded. 
However, the agreement between present numerical results and the experiments is satisfactory. 
In particular, the time evolution of water level on a longer time scale ($t\sqrt{g/H}\approx$ 7--10) 
is evidently in good agreement with experiments in higher spatial resolutions. 
Note that, the presently predicted $h_1$ signals and those obtained by the single-phase
SPH method of Zhang et al. \cite{zhang2017weakly} are in a fair agreement. Nevertheless,
the comparison for $h_2$ reveals the effect of air particles, as the peak value of $h_2$
does not experience an abrupt jump in the present results 
and is thus in a closer agreement with the measurements. 
\subsection{Two-phase liquid sloshing}
Motion of a liquid which partially fills a tank 
due to the periodic motion of the latter is known as liquid sloshing. 
If the motion frequency is close to the natural frequency of the liquid due to gravity wave, 
resonant condition happens, which may cause enormous impact loads on the structures.  
Due to the highly non-linear phenomena in sloshing, 
this benchmark case has been used to validate many single-phase SPH methods 
(see e.g. \cite{rafiee2011study,WINKLER2017165,Zhang_efficientSPH_2018}), 
as well as few two-phase counterparts \cite{mokos2017shifting}. 
In this section we consider a two-phase liquid sloshing case, 
which has been experimentally studied by Rafiee et al. \cite{rafiee2011study}. 

\begin{figure}[tb!]
	\centering
	\includegraphics[width=0.5\linewidth]{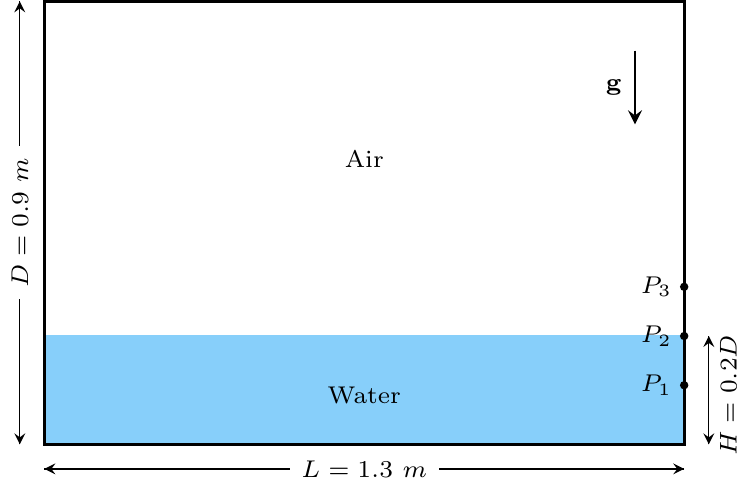}
	\caption{Schematic illustration of the two-phase liquid sloshing problem.}
	\label{fig:2phase_sloshing_config}
\end{figure}

The schematic of this problem is illustrated in Fig. \ref{fig:2phase_sloshing_config}. 
A tank of height $D=0.9$~m and width of $L=1.3$~m 
is partially filled with water to 20\% of the height, i.e. $H=0.18$~m, 
while the remainder is filled by air. 
We consider an inviscid flow, where the density of water and air are set to 
$\rho_w = 1000~\mathrm{kg.m^{-3}}$ and $\rho_a = 1~\mathrm{kg.m^{-3}}$, respectively. 
The gravity is $g=9.81~\mathrm{m.s^{-2}}$ in negative $y$-direction. 
The tank motion is defined by a sinusoidal excitation of $x = A_0sin(2.0 f_0 \pi t)$, 
where $A_0 = 0.1~m$ and $f_0 = 0.496~s^{-1}$ are the amplitude 
and frequency, respectively. 
$f_0$ is close to the natural frequency of water, thereby resulting in resonant condition.
To record the pressure signals, three sensors $P_1$, $P_2$ and $P_3$ are located 
on the right wall at 0.165 m, 0.18 m and 0.195 m from the bottom, respectively.
The same method is utilized to record pressure signals as explained in Section 
\ref{subsec:twophase_dambreak}.

\begin{figure}[tb!]
	\centering
	\includegraphics[width=\linewidth]{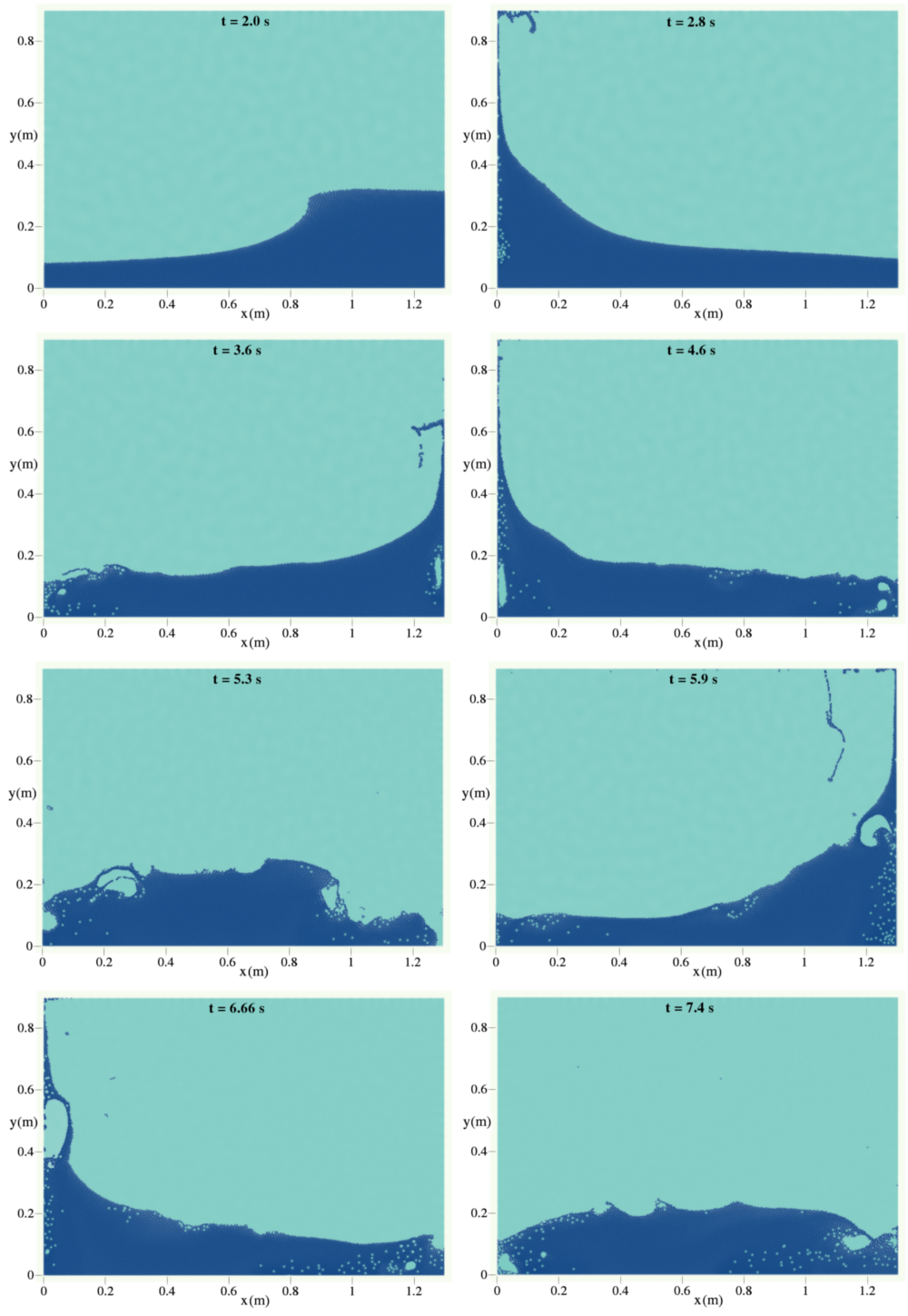}
	\caption{Two-phase liquid sloshing: snapshots of the computational 
		results at eight different time instants 
		$t=$2.0, 2.8, 3.6, 4.6, 5.3, 5.9, 6.66 and 7.4 s with $L/dx=260$.}
	\label{fig:sloshing_8frames}
\end{figure}

\begin{figure}[tb!]
	\centering
	\includegraphics[width=\linewidth]{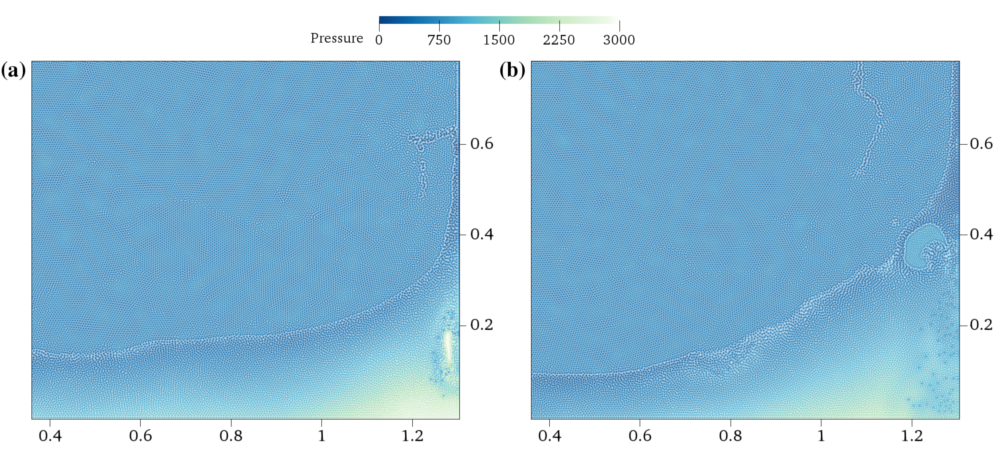}
	\caption{Two-phase liquid sloshing: zoom in view of the free surface breaking, 
		entrapped air pockets and the pressure field at (a) $t=$3.6~s and (b) $t=$5.9~s.}
	\label{fig:sloshing_pressureCloseup}
\end{figure}

Fig. \ref{fig:sloshing_8frames} illustrates several snapshots during the time evolution.  
It can be observed that the highly non-linear features of the flow, viz. strong 
impacts on the walls and largely deformed free surfaces due to re-entry of the backward waves, 
are well captured by the present method, as shown in the experimental and single-phase simulation results 
in Rafiee et al. \cite{rafiee2011study}.
Moreover, the phase interface is sharply maintained between air and water, 
however, air particles are trapped in water to form air bubbles due to highly violent hydrodynamic events. 
Again, evidently no unnatural voids or unrealistic phase separation appears. 
Fig. \ref{fig:sloshing_pressureCloseup} shows the zoom-in view of the water surface 
breaking, entrapped air pockets and the pressure field at two different time instants
from Fig. \ref{fig:sloshing_8frames}.
As it can be observed, the pressure field is quite smooth, while differences in 
its gradient at the phase interface are notable. 
Such near-interface pressure distribution is similar to the results obtained 
in Ref. \cite{chen2015multi} (their Fig. 19) 
where a pressure-gradient discontinuity is imposed at phase interface \cite{hu2007incompressible}. 
Note that, the air pockets entrapped during the impact events are well captured, 
being in agreement with experimental observations of Rafiee et al. \cite{rafiee2011study}. 

\begin{figure}[tb!]
	\centering
\includegraphics[trim = 0mm 0mm 0mm 0mm, clip,width=0.6\textwidth]{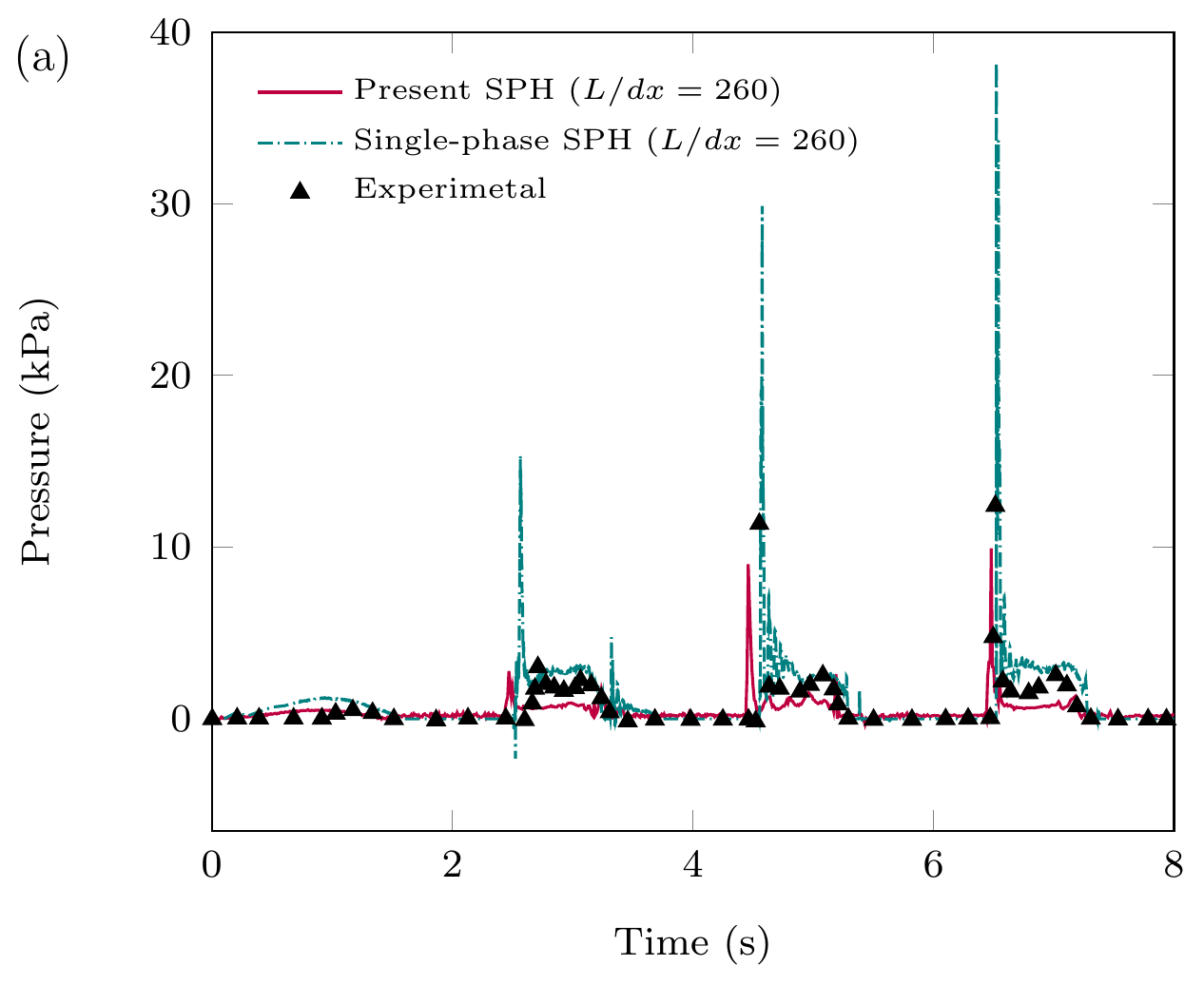}
	\includegraphics[trim = 0mm 0mm 0mm 0mm, clip,width=0.6\textwidth]{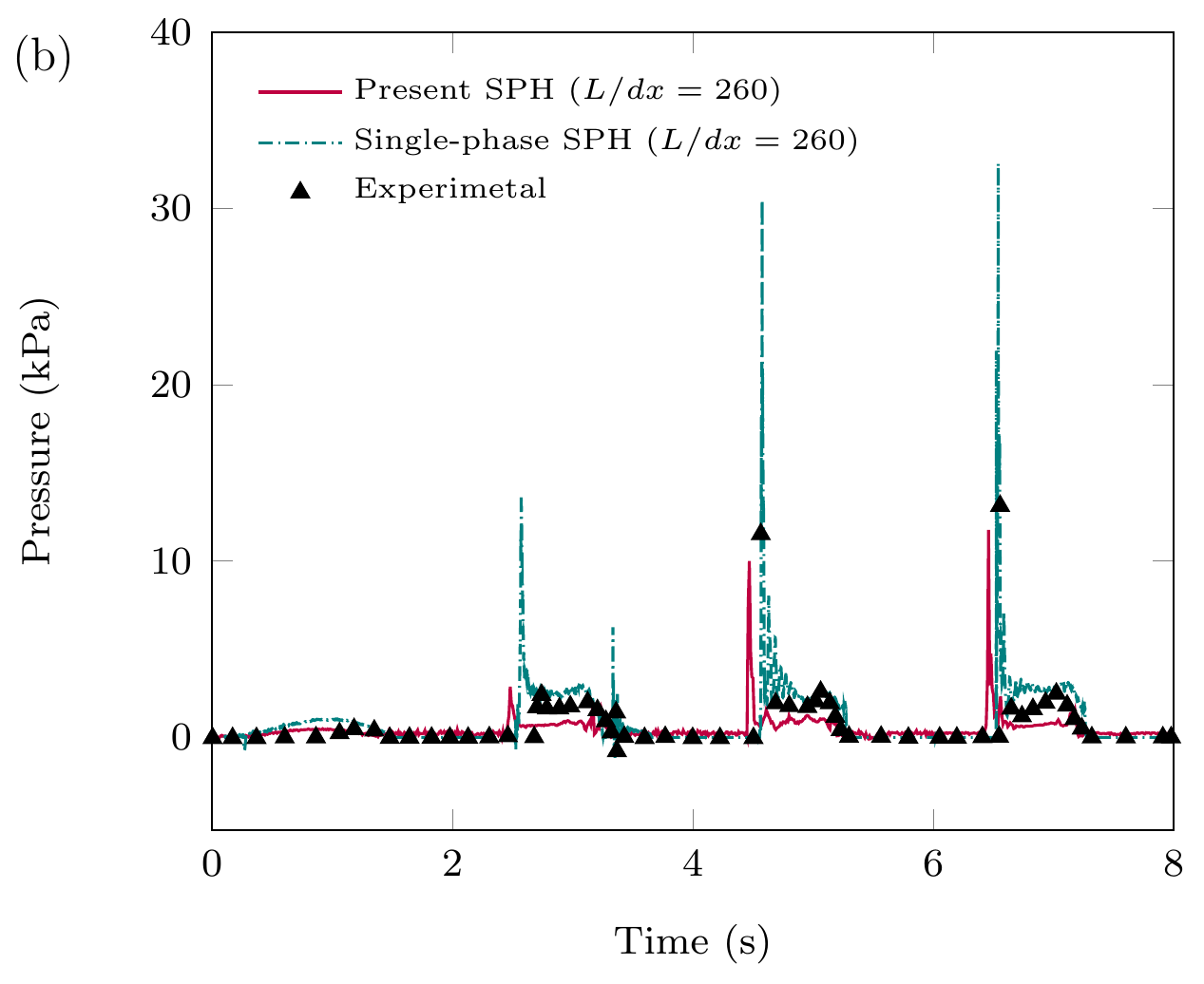}
	\includegraphics[trim = 0mm 0mm 0mm 0mm, clip,width=0.6\textwidth]{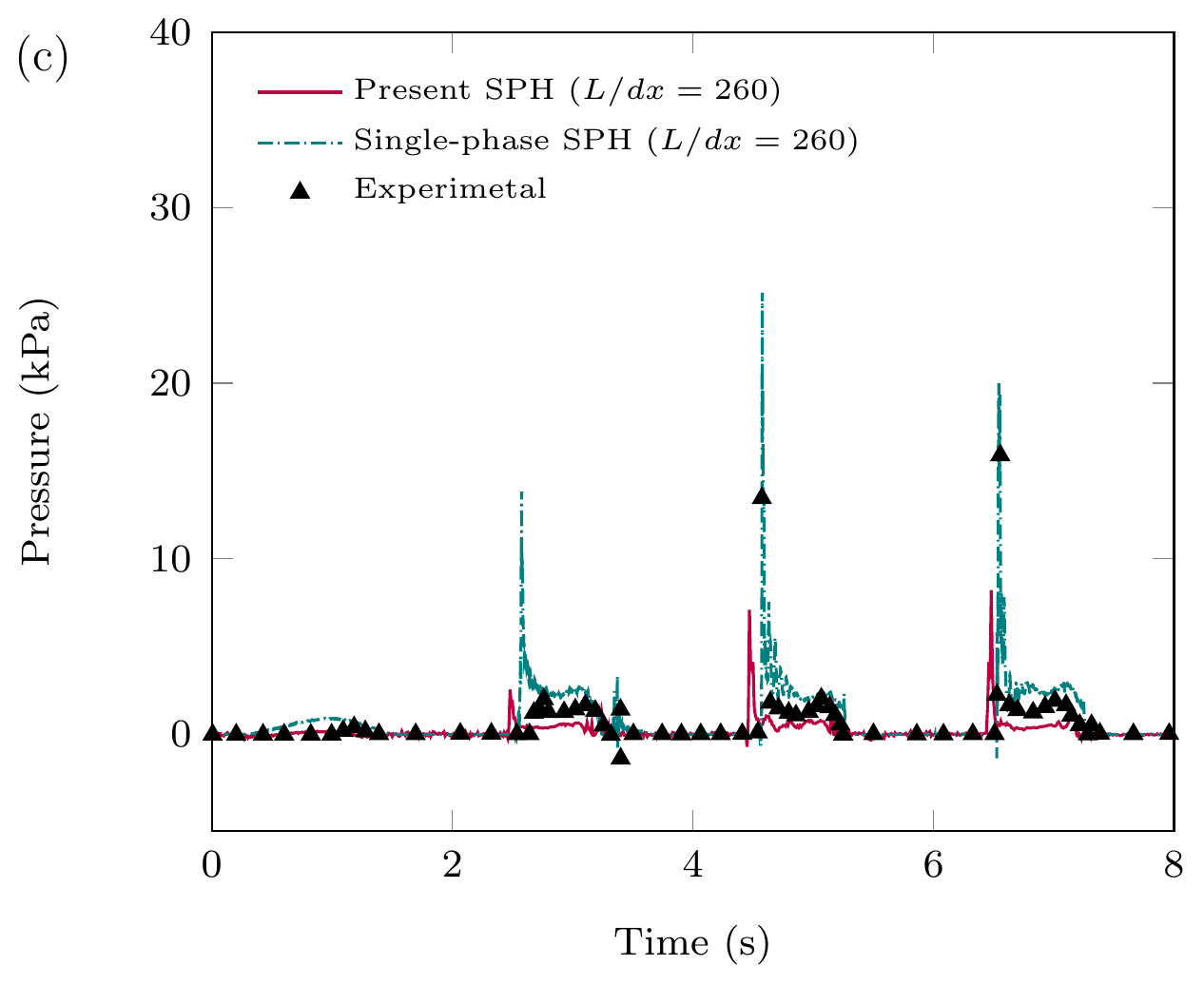}
	\caption{Two-phase liquid sloshing: time history of the impact pressure at sensor 
		(a) $P1$, (b) $P2$ and (c) $P3$ in comparison with experimental data \cite{rafiee2011study} 
		and single-phase SPH predictions presented in \cite{Zhang_efficientSPH_2018}.}
	\label{fig:sloshingpressure}
\end{figure}

As shown in Fig. \ref{fig:sloshingpressure}, 
the time history of the recorded impact pressure signals on the right vertical wall 
are compared against the experimental data \cite{rafiee2011study} 
and the simulation results obtained by a single-phase WCSPH method \cite{zhang2017weakly}.
A good agreement in the overall behavior of the pressure 
is noted in the prediction of both time and value, 
with lower peak pressure values compared to those from experiments. 
Furthermore, in contrast to the single-phase SPH simulation, 
the peak pressure values predicted by the present method are much lower, 
implying the cushioning effect of entrapped air pockets, 
which is excluded in single-phase SPH simulations 
and can lead to highly overestimated peak pressure values \cite{WINKLER2017165}.  

\subsection{3D two-phase dam break}
In order to demonstrate the capabilities of the presented method to simulate large scale problems in 3D, 
the parallel computational framework proposed in Ref. \cite{JI2018} is adopted to carry out 
3D simulations of the two-phase dam break problem. 
The configuration is identical to 
the case in Section \ref{subsec:twophase_dambreak} with the consideration of a third dimension of length $H$. 
To discretize the computational domain, 
particles are initially located on a regular grid with a particle spacing of $dx=H/40$.

\begin{figure}[tb!]
	\centering
	\includegraphics[trim = 0mm 0mm 0mm 0mm, clip,width=0.485\textwidth]{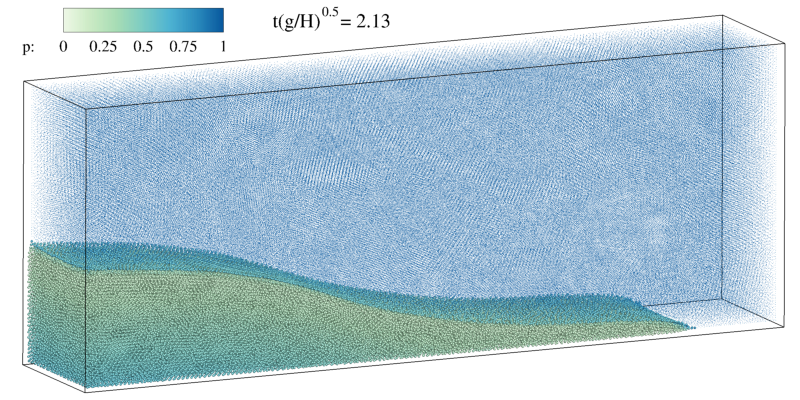}
	\includegraphics[trim = 0mm 0mm 0mm 0mm, clip,width=0.485\textwidth]{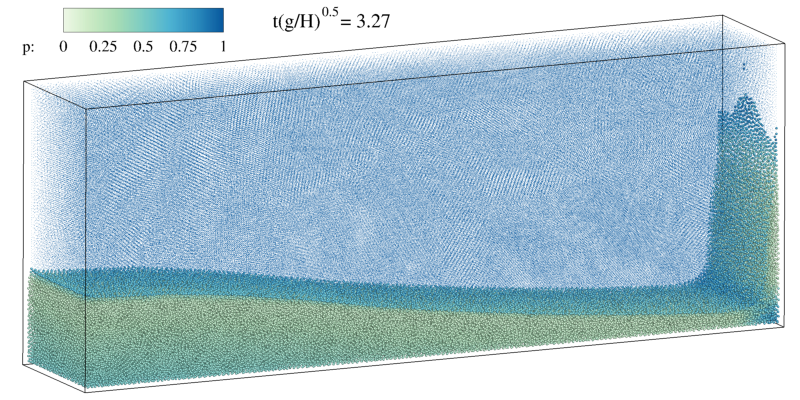}\\
	\includegraphics[trim = 0mm 0mm 0mm 0mm, clip,width=0.485\textwidth]{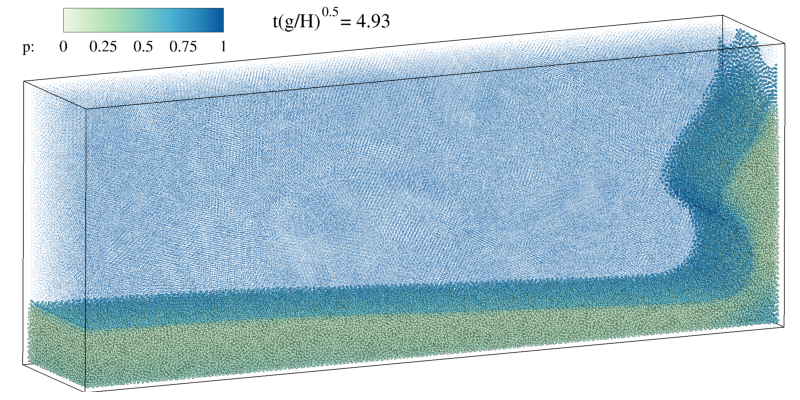}
	\includegraphics[trim = 0mm 0mm 0mm 0mm, clip,width=0.485\textwidth]{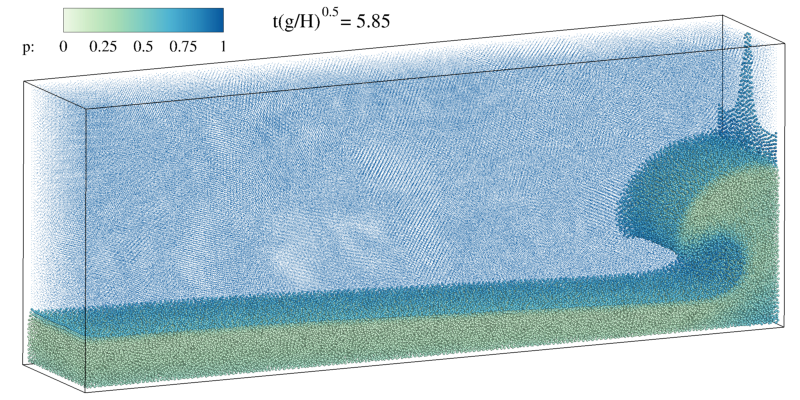}\\
	\includegraphics[trim = 0mm 0mm 0mm 0mm, clip,width=0.485\textwidth]{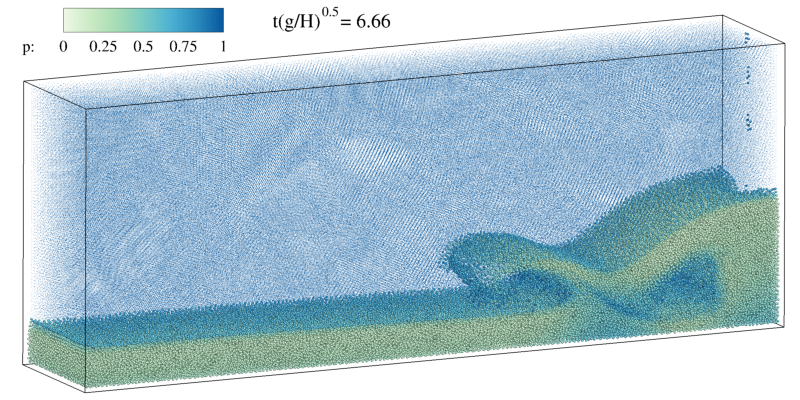}
	\includegraphics[trim = 0mm 0mm 0mm 0mm, clip,width=0.485\textwidth]{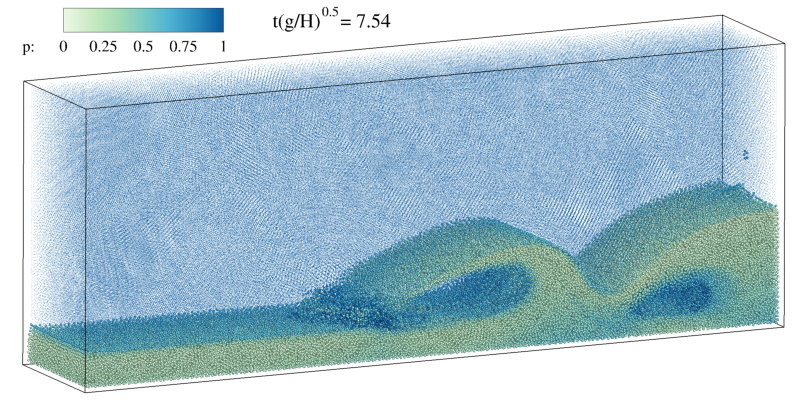}\\
	\includegraphics[trim = 0mm 0mm 0mm 0mm, clip,width=0.485\textwidth]{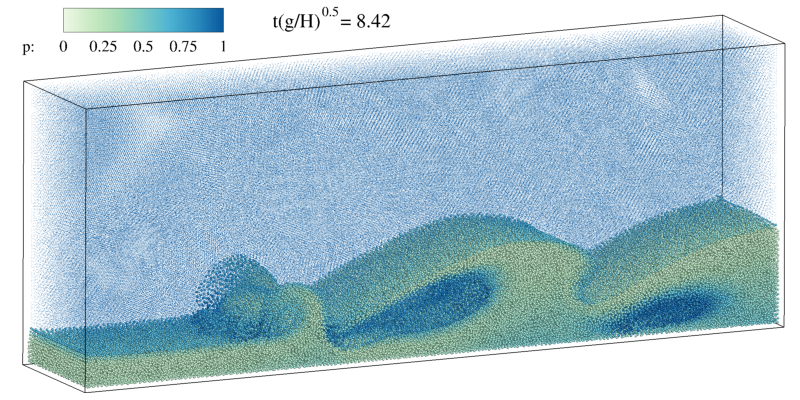}
	\includegraphics[trim = 0mm 0mm 0mm 0mm, clip,width=0.485\textwidth]{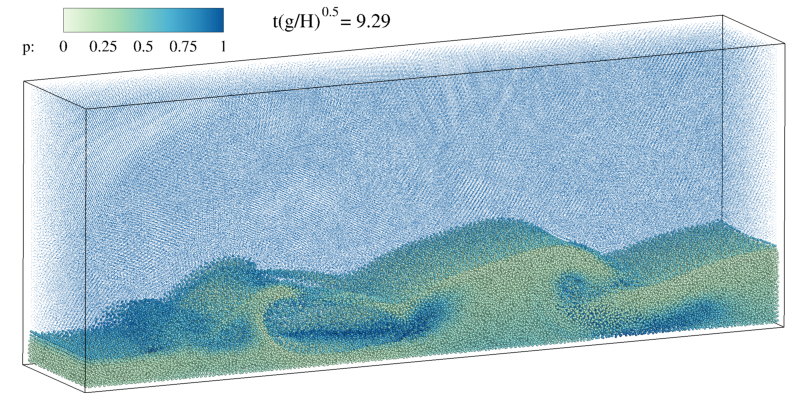}
	\caption{3D two-phase dam break: snapshots of particle and pressure distribution. }
	\label{figs:damfreesurface}
\end{figure}

Fig. \ref{figs:damfreesurface} illustrates several snapshots during the time evolution of the free surface. 
Similar to the presented results in Section \ref{subsec:twophase_dambreak}, 
the main features of the dam break problem are well captured also here.  
Evidently, as shown in Fig. \ref{figs:damfreesurface}, 
the pressure field is quite smooth, 
while large gradient is observed near the phase interface. 
Again, the entrapped air pockets are well captured during the impact events 
and the phase interface is sharply maintained with no spurious fragmentation, 
and no particle penetration is observed.

\section{Conclusions}\label{sec:conclusions}
In this paper, we have presented a multi-phase SPH method for 
the simulation of highly violent flows with large density ratio.
With the particle interactions being handled by a multi-phase Riemann solver, 
the scheme is simple and robust. 
The method employs the same values for the artificial speed of sound in both heavy and light phases. 
This advantage, allows for larger time steps, 
which greatly improves the computational performance.
The proposed method is validated against existing experimental data, 
previous numerical studies and analytical solutions for several water-air flow problems. 
It is observed that stable solutions with sharp interfaces are obtained
without suffering from non-physical void regions which is observed in previous SPH simulations, 
especially in those using the same choice of artificial sound speeds. 
It is also observed that the accuracy of the present method is
comparable to the previous less efficient methods using a much larger speed of sound in the light phase. 
Furthermore, the present method replicates the cushioning effect of entrapped air pockets, 
leading to more reasonable pressure peaks. 
Albeit the present work focuses on benchmark test cases, 
the method is sufficiently generic to be employed in more complex scientific 
and industrial problems, which is the main scope of future work.
\section*{Acknowledgments}
The authors gratefully acknowledge the financial support by Deutsche 
Forschungsgemeinschaft (DFG HU1572/10-1, DFG HU1527/12-1) for the present work.

\section*{References}

\bibliography{bibliography}

\end{document}